\documentclass[prd,twocolumn,superscriptaddress,longbibliography,preprintnumbers,floatfix,nofootinbib,10pt,aps]{revtex4-1}
\usepackage[utf8]{inputenc}
\usepackage[T1]{fontenc}
\usepackage{url}
\usepackage{xspace}
\usepackage{dsfont}
\usepackage{amssymb}
\usepackage{amsmath}
\usepackage{graphicx}
\usepackage[colorlinks=true,citecolor=blue]{hyperref}
\usepackage{xcolor}
\usepackage{subcaption}
\usepackage{multirow}
\usepackage{hhline}
\usepackage{comment}
\usepackage{booktabs}

\begin{document}

\title{An AI-ready, Polarized Electron-Positron Collision Dataset}

\author{Chi Lung Cheng}
\email{alkaidc@stanford.edu}
\affiliation{SLAC National Accelerator Laboratory, Menlo Park, CA 94025, USA}

\author{Simon Corrodi}
\email{scorrodi@anl.gov}
\affiliation{High Energy Physics Division, Argonne National Laboratory, Lemont, IL 60439, USA}

\author{T. J. Hobbs}
\email{tim@anl.gov}
\affiliation{High Energy Physics Division, Argonne National Laboratory, Lemont, IL 60439, USA}

\author{Alaettin Serhan Mete}
\email{amete@anl.gov}
\affiliation{High Energy Physics Division, Argonne National Laboratory, Lemont, IL 60439, USA}

\author{Benjamin Nachman}
\email{nachman@stanford.edu}
\affiliation{SLAC National Accelerator Laboratory, Menlo Park, CA 94025, USA}

\begin{abstract}
We present a modernized, AI-ready release of reconstructed data from the SLD experiment at the SLAC Linear Collider (SLC).  The dataset comprises approximately 660{,}000 reconstructed events collected at $\sqrt{s}\approx 91.2$~GeV with a highly polarized electron beam from 1996--1998.  The data have been translated from legacy formats into modern, widely-used file formats with the help of AI agents. The release also includes a corpus of newly digitized SLD internal documentation. We describe the contents of both components and provide physics validation demonstrations along with illustrations of their utility for physics and machine learning research in particle physics.
\end{abstract}

\maketitle

\section{Introduction}

\begin{figure}
    \centering
    \includegraphics[width=0.95\linewidth]{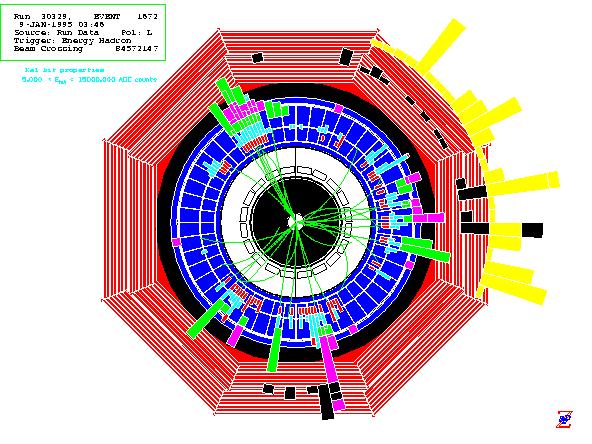}
    \caption{ An event display of a collision in the SLD detector. Reproduced from \href{https://stanford.edu/~dubois/stained_glass/projects/sldstep/event.jpg}{this website}.}
    \label{fig:eventdisplay}
\end{figure}

The SLAC Linear Collider (SLC)~\cite{SLC:1980report} was the first and only high-energy linear $e^+e^-$ collider, operating at the $Z$ pole ($\sqrt{s}\approx 91.2$~GeV) from 1989 to 1998.  Its unique feature was a highly polarized electron beam, achieving longitudinal polarizations of $\sim$73--77\%~\cite{SLD:2000leq}, which enabled precision electroweak measurements not accessible at unpolarized colliders.  The SLD experiment~\cite{SLD:1984design} ran from 1992 to 1998 at the $Z$ pole. The high-polarization 1996--1998 portion, covered by the present release, alone comprises approximately 660{,}000 reconstructed events.
SLD made pioneering contributions to heavy flavor physics and electroweak precision tests, including the world's most precise single measurement of the weak mixing angle $\sin^2\theta_W^{\text{eff}}$ via the left-right asymmetry $A_{LR}$~\cite{SLD:2000leq}.  The detector featured an innovative CCD-based vertex detector (VXD)~\cite{SLD:1996vxd3}, which provided exceptional impact parameter resolution and enabled high-purity flavor tagging of $b$- and $c$-quark jets.
A representative reconstructed $Z$ decay in the SLD detector is shown in Figure~\ref{fig:eventdisplay}.

Despite the lasting physics impact of SLD, the reconstructed data have remained in legacy formats that are inaccessible to modern analysis tools.  As interest in applying machine learning (ML) to particle physics continues to grow~\cite{Carleo:2019ptp}, there is significant value in making diverse experimental datasets available in standardized, ML-ready formats.  Electron-positron collision data are particularly attractive for ML development because the initial state is precisely known, the events are relatively clean compared to hadron collisions, and the well-understood theoretical predictions provide reliable ground truth for benchmarking.
Equally important is the institutional knowledge that enables analysts to interpret legacy datasets, including data format conventions, calibration choices, and selection rationales.  For older experiments, much of this lives only in internal notes that were never born digital.

The diversity of public datasets available for training and benchmarking is itself a first-order bottleneck for ML in particle physics.  In collider physics, the vast majority of currently public datasets come from a single type of collision---proton-proton ($pp$) collisions at the LHC---leaving the $e^+e^-$ and $ep$ regimes underrepresented.  Reviving high-quality legacy data from these other regimes broadens the range of physics conditions available for downstream analyses and applications, and the SLD is a uniquely valuable target because the longitudinal polarization of its initial state is a configuration that has not been realized at any other collider.  SLD therefore remains the only $Z$-factory dataset where the helicity of the initial state is known event by event.

In this paper, we describe the modernization and public release of the SLD reconstructed dataset, including its translation into modern, AI-ready columnar formats. The SLD Collaboration software, written in an extension of the Fortran programming language (Mortran), relied on a proprietary and poorly documented ecosystem that is now largely unavailable, posing a significant challenge for accessing and interpreting the data. 
In this context, AI-assisted tools were used to aid in the interpretation and recovery of previously undocumented or partially documented data structures. This provides a representative example of how legacy datasets with incomplete software or documentation can be made accessible using modern tools. The release is accompanied by digitized internal documentation previously available only in physical form, along with text extracted by several state-of-the-art document-processing tools to make the corpus directly usable in modern AI workflows. We also showcase the utility of the dataset for modern ML research by passing it through a particle physics foundation model, and demonstrate the AI-readiness of the documentation release through an agentic question-answering system.

\section{Dataset} \label{sec:dataset}

The released dataset is hosted on Zenodo~\cite{Cheng:2026zenodo} and consists of two components: the reconstructed collision data covering the 1996--1998 SLD runs, and the digitized internal SLD documentation. The 1992--1995 data have not yet been recovered and are not part of this release.

\subsection{Reconstructed Data}

After data taking concluded, the analysis software that decoded the binary banks fell into disuse.  The original developers had dispersed, no public specification of the SLD data binary format (Jazelle) bank layout existed outside of internal notes (some recoverable only through the Internet Archive), and none of the original Mortran reader code remains operational on modern systems.  We were fortunate to acquire a partial Java reader written in 2021 by the former SLD software coordinator.  While highly useful, this reader covers only a subset of the reconstructed bank families and does not interoperate with contemporary columnar formats or array libraries.  For example, it does not expose the \texttt{PHBM} beam-information bank, which records the per-event electron-beam polarization and its associated uncertainty.  Without this bank, none of SLD's polarized observables, including the left-right asymmetry $A_{LR}$, can be reconstructed. The reconstructed binaries themselves were preserved on archival storage, but without a complete and modern reader the dataset was inaccessible to current analysis tools.

We reconstructed the missing bank layouts by reverse-engineering the binary format with AI assistance (Anthropic's Claude).
Two sources of information guided the recovery.  The first is partial documentation of bank contents preserved in the recovered SLD internal notes, which specifies which fields each bank should carry but is typically silent on, or outdated about, how those fields are serialized and compressed on disk.  The second is physics-based ground truth: the kinematics of well-understood processes such as $Z\to q\bar{q}$ fix the numerical values that must appear in each event, providing a strict test for any proposed layout.  Working from this combination, we identified candidate field positions and data types from recurring byte patterns and alignments across event dumps.  Candidates that failed to reproduce the expected values across additional events were refined in the next iteration, with the cycle typically converging for each bank family after several rounds.  In this way we decoded the principal MiniDST bank families, including several not covered by any surviving implementation, and packaged the result into the \texttt{jazelle} reader described below.  A small number of bank families remain incompletely decoded.  These are well-defined residual problems expected to be tractable with the same approach given additional test events or documentation.  Beyond making SLD data usable again, this exercise illustrates a broader pattern: data files that outlive the software written to decode them are now recoverable by combining surviving documentation with physics-based constraints and modern AI assistance.

The translation pipeline is released as the open-source Python package \texttt{jazelle}~\cite{Cheng:jazelle_reader}.  It reads the legacy Jazelle binary format and emits Awkward~\cite{Pivarski:awkward} record arrays that can be serialized to several AI-friendly columnar formats including HDF5, Parquet, and Feather.  The public Zenodo release uses Parquet as a convenient default but can be regenerated in any of the supported formats by running the toolkit on the preserved binaries that ship alongside the columnar release.  The translation preserves the original SLD bank hierarchy, so that the released arrays remain navigable for users familiar with the legacy layout and every quantity available to the original collaboration remains accessible.

Each event includes the following information:
\begin{itemize}
\item \textbf{Event header:} run and event numbers, timestamp, and trigger information.
\item \textbf{Beam information:} center-of-mass energy and per-event electron-beam polarization with its associated uncertainty.
\item \textbf{Charged tracks:} helix parameters and their covariance matrix, hit counts, fit quality, and $dE/dx$ from the Central Drift Chamber and the VXD3 vertex detector~\cite{SLD:1996vxd3}.
\item \textbf{Calorimeter clusters:} raw and per-layer energy depositions from the Liquid Argon Calorimeter (LAC).
\item \textbf{Particle-identification subsystems:} Cherenkov ring likelihoods from the Cherenkov Ring Imaging Detector (CRID), muon-track segments from the Warm Iron Calorimeter (WIC), and electron-calorimeter matching information.
\item \textbf{Relational tables:} cross-references that link reconstructed objects across subsystems, for example associating tracks with calorimeter clusters.
\end{itemize}
The full set of mini Data Summary Tape (mini-DST) bank families is preserved in the release.  The items above highlight the principal ones used in the validation studies below.

The release applies only the standard SLD reconstruction-level data-quality requirements that suppress non-collision backgrounds such as beam-gas interactions and two-photon processes.  No channel-specific event selection is pre-applied at the release stage.  The published selections used for the validation studies are described in Sec.~\ref{sec:physics_validation}.

\subsection{Documentation}

A collection of internal SLD notes and technical documents have been digitized and are included alongside the data release.  These documents, previously available only as paper copies in the SLAC archives, describe the detector subsystems, calibration procedures, reconstruction algorithms, and analysis techniques used by the SLD collaboration.  They provide essential context for understanding the dataset and its associated systematic effects.

The structure of the documentation release and its supporting evaluation are described separately in Sec.~\ref{sec:documentation}.

\section{Physics Validation and Measurement Baselines} \label{sec:physics_validation}

As a demonstration of the dataset's capability for physics analysis, we reproduce two canonical SLD analyses on the released 1996--1998 data: the hadronic left-right cross-section asymmetry $A_{LR}$ and the leptonic coupling asymmetries $A_\ell$.  For the hadronic channel we adopt the event selection of the 2000 $A_{LR}$ analysis~\cite{SLD:2000leq}, which is the latest published SLD measurement using the 1996--1998 dataset.  For the leptonic channels we use the selections of the 2001 leptonic-coupling analysis~\cite{SLD:2000ujp}, the latest SLD leptonic-coupling measurement using the same year-range.  Both selections are applied to the released data without modification.  Following the original analyses, we pool the 1997 and 1998 running periods into a single year-group because the SLC polarized source and beam operating point were essentially unchanged across those two years, and report 1996 separately.  All results in this section are intended as a demonstration of internal consistency rather than as an actual measurement, and uncertainties quoted include only statistical and beam-polarization contributions.

\subsection{Kinematic distributions}

A first check is that the kinematic distributions reconstructed from the translated data match the well-understood physics of $Z$ decays at the $Z$ pole.

\begin{figure*}[t]
\centering
\begin{subfigure}[t]{0.42\textwidth}
\includegraphics[width=\linewidth]{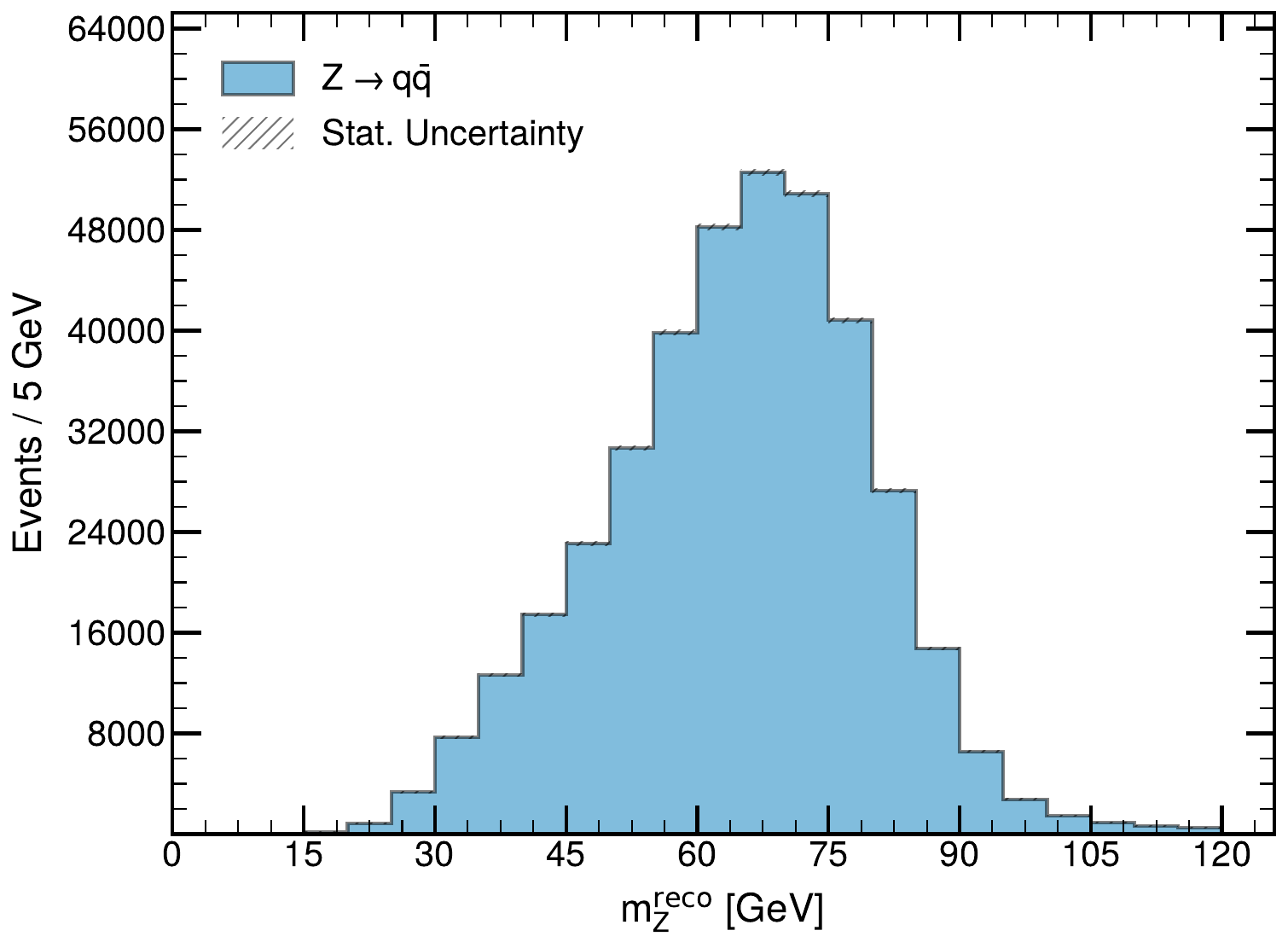}
\caption{Hadronic $Z\rightarrow q\bar{q}$.}
\end{subfigure}\hfill
\begin{subfigure}[t]{0.42\textwidth}
\includegraphics[width=\linewidth]{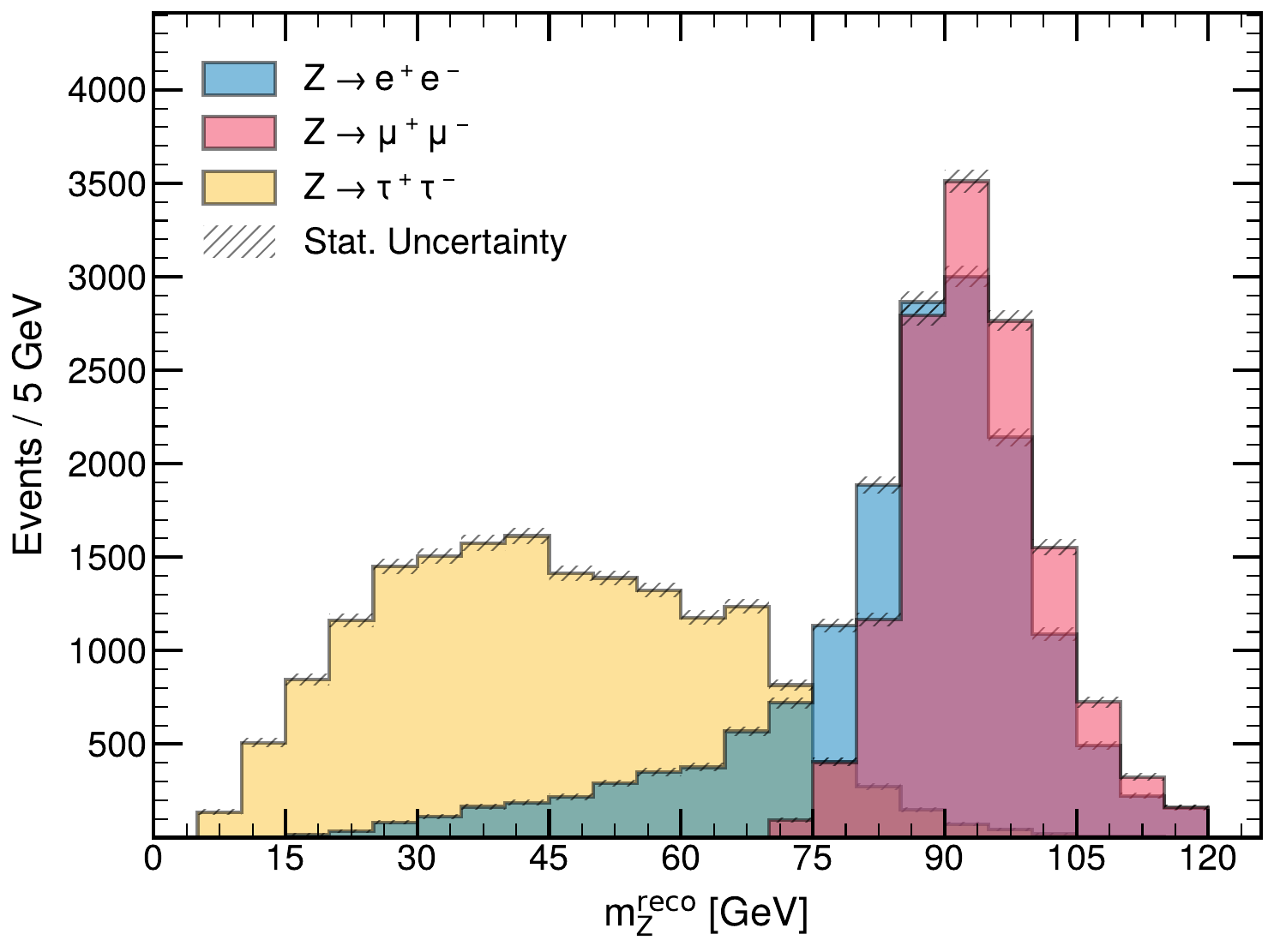}
\caption{Leptonic $Z\rightarrow \ell\bar{\ell}$.}
\end{subfigure}
\caption{Reconstructed visible invariant-mass spectra: (a) hadronic $Z\!\to\!q\bar{q}$ channel, (b) the three leptonic $Z$ decay channels overlaid.}
\label{fig:mass}
\end{figure*}

Figure~\ref{fig:mass} shows the reconstructed visible-mass spectra for the four selections.  The hadronic spectrum peaks $\sim\!20$~GeV below $m_Z^{\mathrm{reco}}$ because of energy lost to long-lived neutral hadrons in the calorimeter and to neutrinos from semileptonic heavy-flavor decays.  The $e^+e^-$ and $\mu^+\mu^-$ channels reconstruct $m_Z^{\mathrm{reco}}$ accurately as expected for fully visible final states, with a low-energy tail in $e^+e^-$ from initial-state radiation.  The $\tau^+\tau^-$ spectrum is broad and smeared, peaking around 40--50~GeV due to energy lost to neutrinos in $\tau$ decays.

\begin{figure}[t]
\centering
\includegraphics[width=0.85\linewidth]{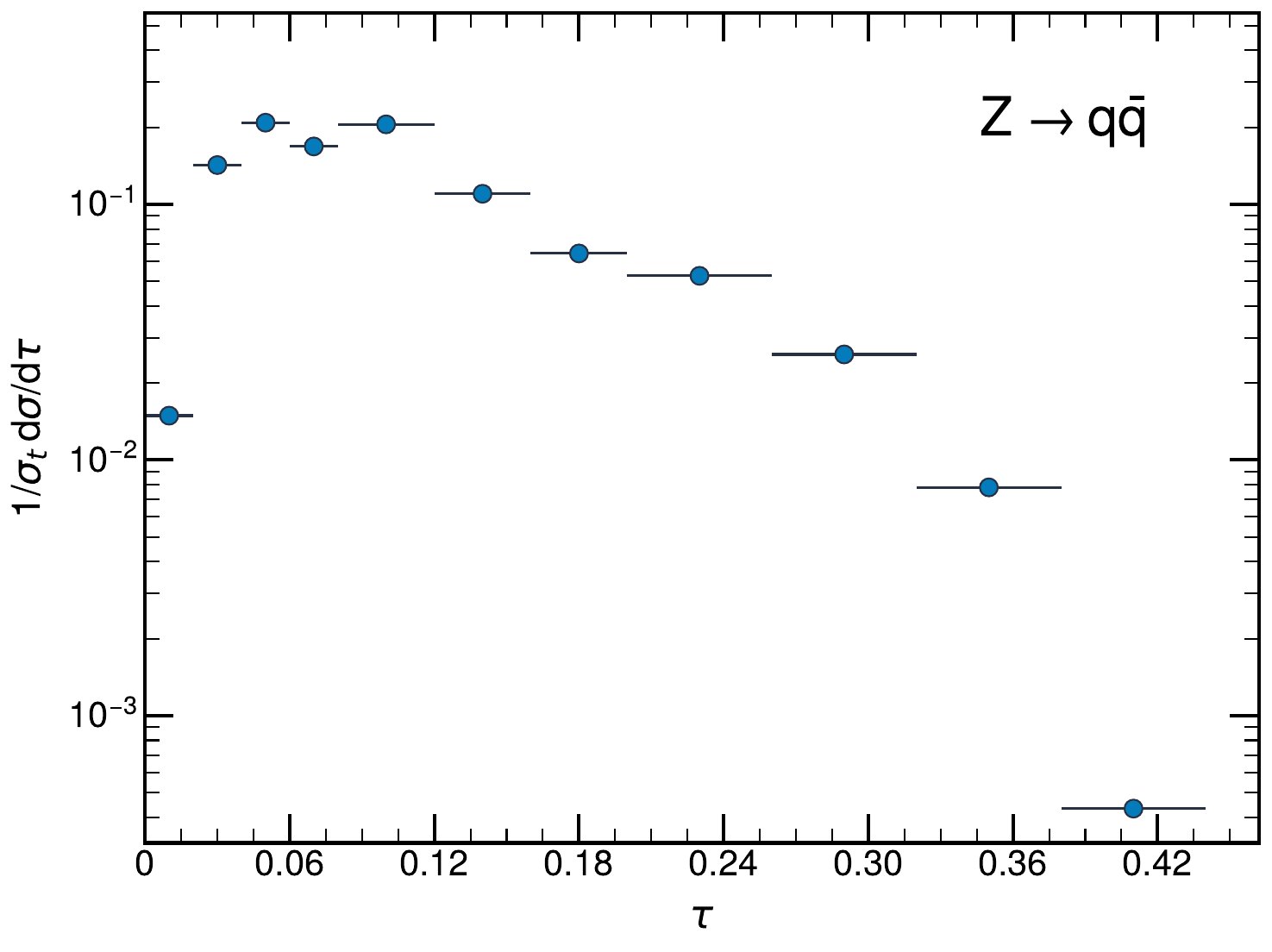}
\caption{Distribution of the event-shape variable $\tau \equiv 1-T$, where $T$ is the thrust, in selected hadronic events.  The peaking of the distribution at small $\tau$ confirms the back-to-back two-jet topology characteristic of $Z\to q\bar{q}$ decays at the $Z$ pole.}
\label{fig:thrust}
\end{figure}

The hadronic event topology can be further examined through the event-shape variable $\tau\equiv 1-T$, where $T$ is the thrust~\cite{Farhi:1977sg}.  Figure~\ref{fig:thrust} shows the $\tau$ distribution for selected hadronic events, peaking at small $\tau$ as expected for back-to-back two-jet final states with relatively rare hard gluon emission populating the tail.  This is the characteristic $Z\to q\bar{q}$ topology at the $Z$ pole and is one of the cleanest checks that the reconstructed final-state objects in the released data are correctly assembled.

\begin{figure*}[t]
\centering
\begin{subfigure}[t]{0.32\textwidth}
\includegraphics[width=\linewidth]{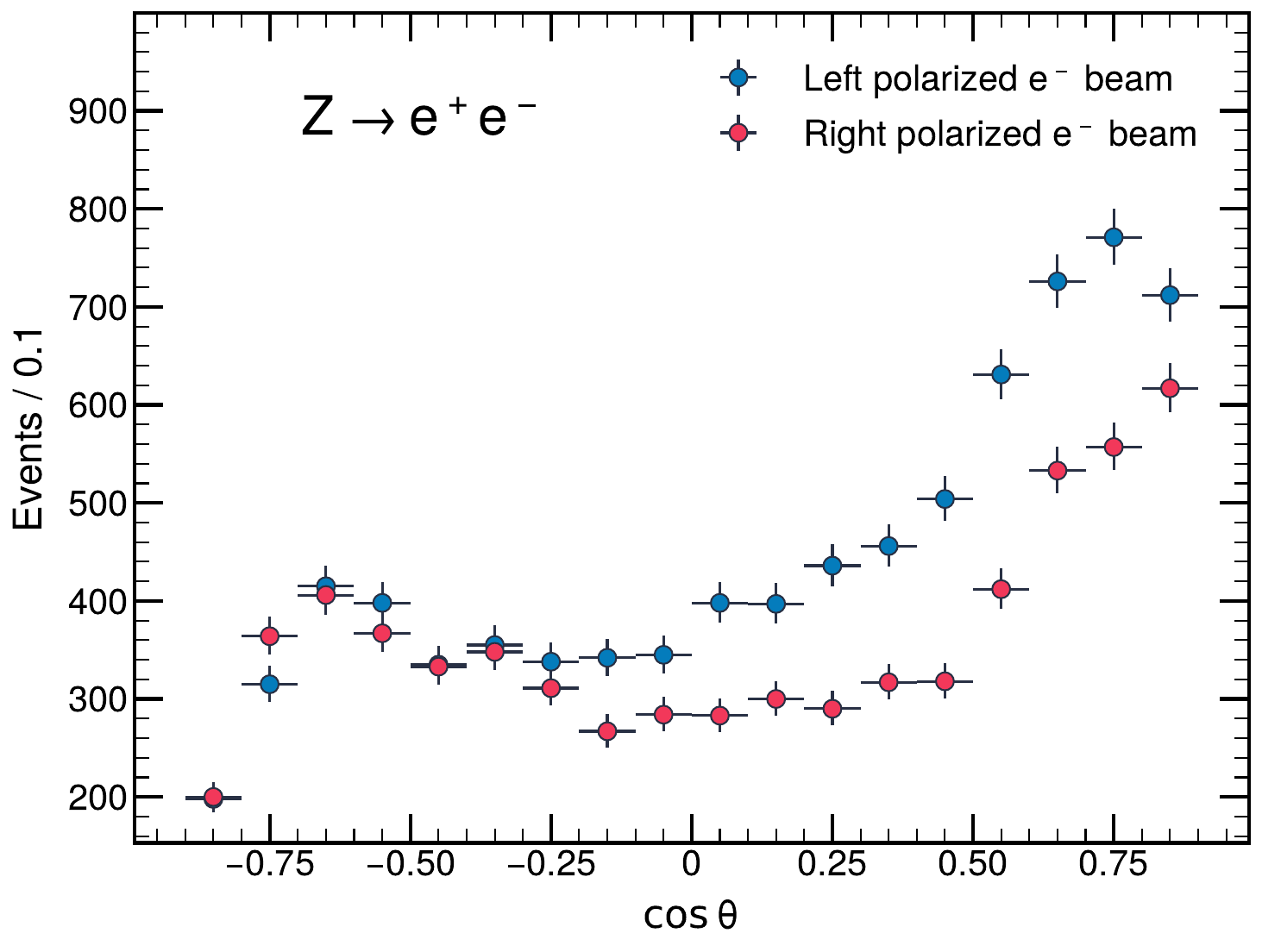}
\caption{$\cos\theta$, $Z\!\to\!e^+e^-$.}
\end{subfigure}\hfill
\begin{subfigure}[t]{0.32\textwidth}
\includegraphics[width=\linewidth]{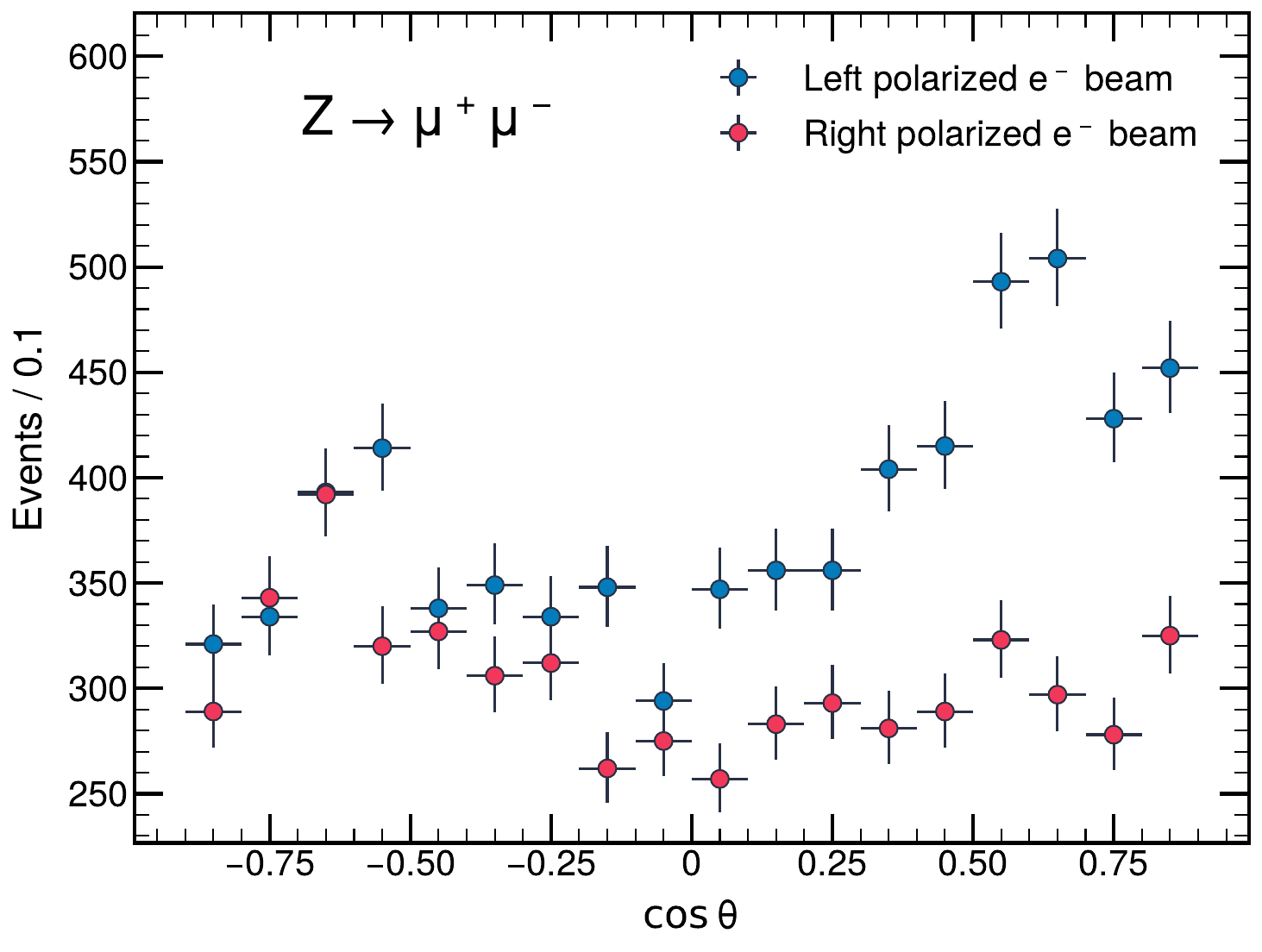}
\caption{$\cos\theta$, $Z\!\to\!\mu^+\mu^-$.}
\end{subfigure}\hfill
\begin{subfigure}[t]{0.32\textwidth}
\includegraphics[width=\linewidth]{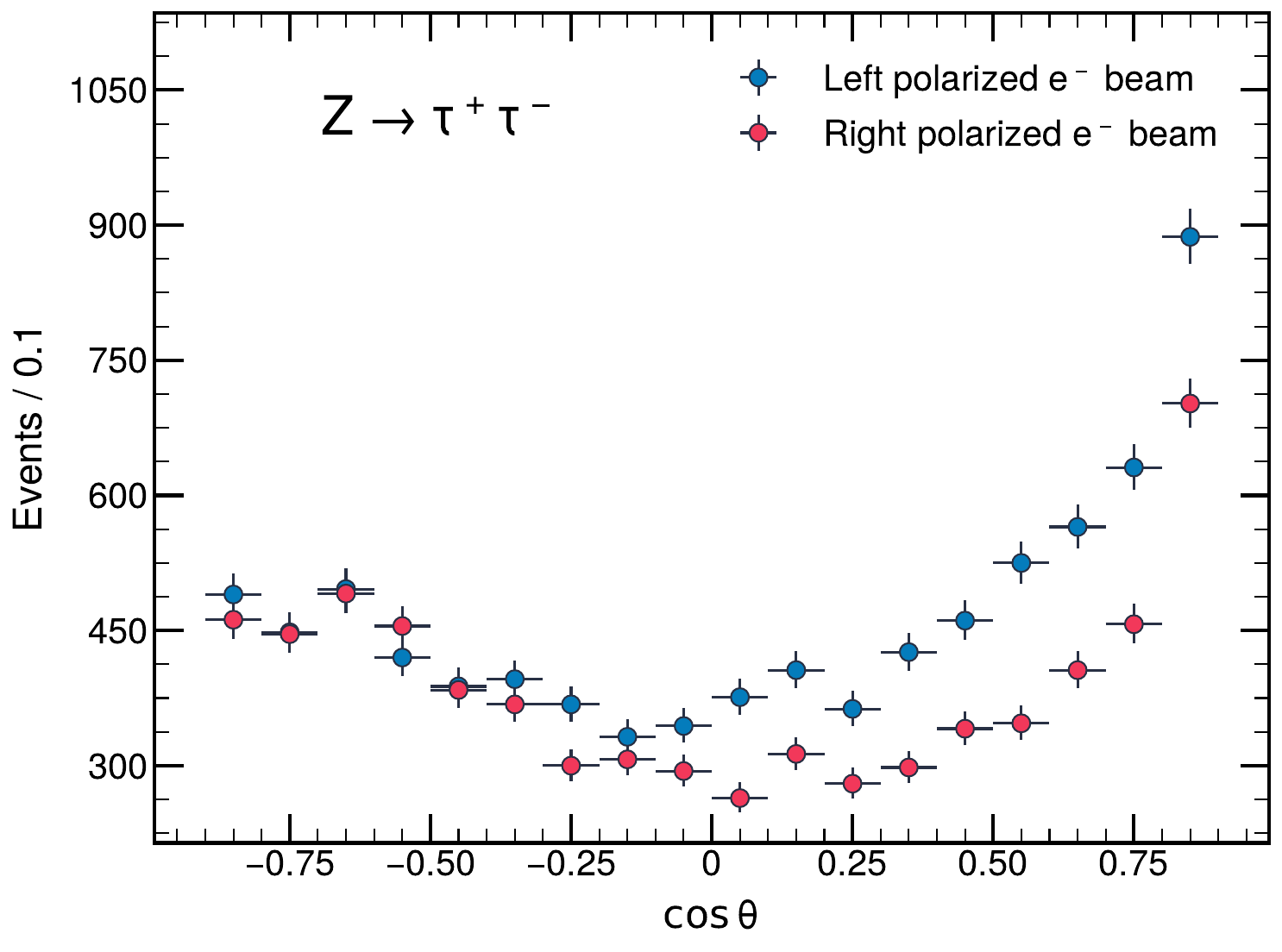}
\caption{$\cos\theta$, $Z\!\to\!\tau^+\tau^-$.}
\end{subfigure}
\caption{Polar-angle distributions for the three leptonic channels in the 1997--1998 subset, separated by the sign of the electron-beam polarization.  The clear $L/R$ splitting is the kinematic origin of the leptonic coupling asymmetries $A_\ell$.}
\label{fig:kin}
\end{figure*}

Figure~\ref{fig:kin} shows the polar-angle distributions of the dilepton final states for the two beam helicities.  At the $Z$ pole, the differential cross-section for $e^+e^-\!\to\!Z\!\to\!\ell^+\ell^-$ has a polarization-dependent forward-backward component proportional to the product of the initial- and final-state coupling asymmetries $\mathcal{A}_e\mathcal{A}_\ell$, which provides the kinematic handle for extracting $\mathcal{A}_\ell$~\cite{SLD:2000ujp}.  Under the assumption of lepton universality ($\mathcal{A}_e=\mathcal{A}_\ell$), the left- and right-polarized distributions are predicted to cross at $\cos\theta=-1$, a feature that is clearly visible in the data.  The steep forward rise in the $ee$ channel comes from the $t$-channel Bhabha contribution that sits on top of the $s$-channel $Z$ exchange.

\subsection{Asymmetry measurements}
SLD was uniquely capable of precise cross-section measurements for the $Z$-pole process $e^+e^- \to Z \to f \bar{f}$ with explicit dependence on the electron-beam polarization, $P_e$, and polar angle of the outgoing fermion, $\theta$. Singly differential cross sections possess the generic form
\begin{equation}
\begin{split}
\frac{d\sigma}{d\cos\theta} \;\propto\;\;
 (1&+\cos^2\theta)\,\bigl[\,1 - P_e\,A_e\,\bigr] \\
& {}+\; 2\cos\theta\,\bigl(A_e - P_e\bigr)\,A_f .
\end{split}
\label{eq:dsigma-polarized}
\end{equation}
As such, detailed empirical information on the cross section of Eq.~(\ref{eq:dsigma-polarized}) covering $\theta$ at distinct $P_e$ furnish an array of forward-backward and left-right polarization asymmetries. These asymmetries in turn depend on electroweak (EW) couplings and parameters which may constrain extractions of potential beyond Standard Model (BSM) physics signatures.
In particular, the fermion-level asymmetries appearing in Eq.~(\ref{eq:dsigma-polarized}) can be written explicitly in terms of the EW couplings,
\begin{equation}
A_f \;=\; \frac{2\, g_V^{f}\, g_A^{f}}{(g_V^{f})^2 + (g_A^{f})^2} \, ,
\label{eq:Af-def}
\end{equation}
where (at tree level) in terms of charge and isospin,
$g_V^{f} = T_3^{f} - 2 Q_f \sin^2\theta_W^{\rm eff}$ and $g_A^{f} = T_3^{f}$, such that the electron asymmetry is
\begin{equation}
A_e \;=\; \frac{2\, v_e}{1 + v_e^2} \, ,
\qquad
v_e \;\equiv\; 1 - 4\sin^2\theta_W^{\rm eff} \, .
\label{eq:Ae-vs-sin2theta}
\end{equation}
Similar measurements and expressions allow joint extractions of $\sin^2\theta_W^{\text{eff}}$, fermion-level asymmetries that test lepton universality, forward-backward asymmetries with final-state flavor tags, and parametrizations of physics beyond the Standard Model, such as within effective field theory. 

From the selected samples we extract the left-right cross-section asymmetry $A_{LR}$ from the hadronic channel and the leptonic coupling asymmetries $A_\ell$ from the leptonic channels by direct event counting, computed per year-group and combined by inverse-variance weighting.  The mean electron-beam polarizations are $\langle\mathcal{P}_e\rangle = (76.14\pm1.42)\%$ for 1996 and $(73.08\pm1.72)\%$ for 1997--1998, the step between the two year-groups reflecting the upgrade of the SLC polarized electron source after 1996.  Quoted uncertainties include statistical and beam-polarization contributions only, with no other systematic uncertainties accounted for since these baselines are intended as a demonstration rather than as an actual measurement.  Selected event counts and extracted asymmetries are summarized in Table~\ref{tab:results}.  The hadronic $A_{LR}$ translates into an effective weak mixing angle $\sin^2\theta_W^{\mathrm{eff}}=0.23144\pm 0.00044$, and the leptonic-channel determination yields $\sin^2\theta_W^{\mathrm{lep}}=0.23278\pm 0.00102$.

\begin{figure}[ht]
\centering
\includegraphics[width=0.95\linewidth]{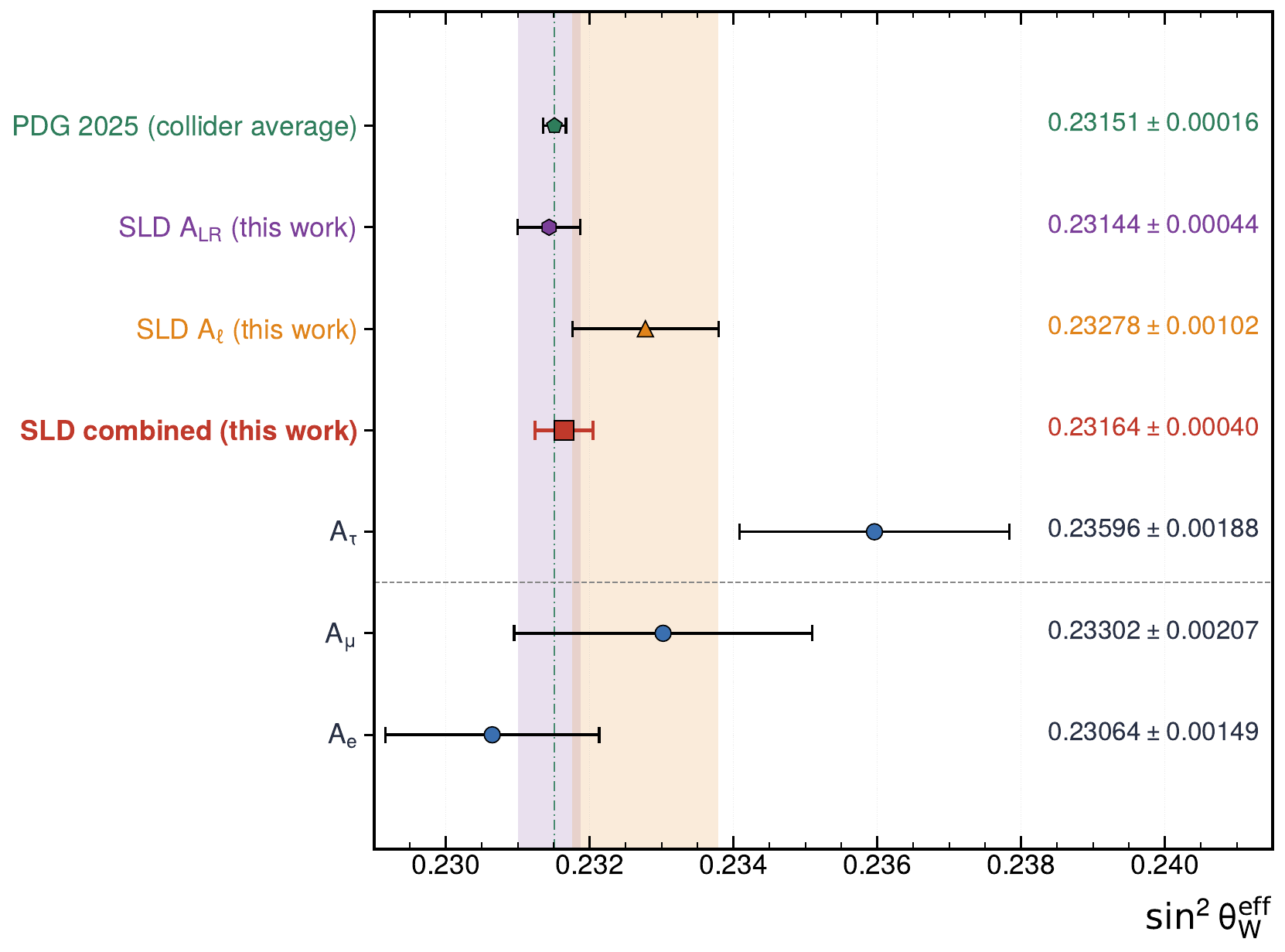}
\caption{Extracted values for the effective weak mixing, $\sin^2\theta_W^{\mathrm{eff}}$, as obtained from $A_{LR}$ and $A_\ell$. The top row shows the collider averaged value from PDG~\cite{ParticleDataGroup:2024cfk}.}
\label{fig:sin2w}
\end{figure}

We illustrate the constraints on $\sin^2\theta_W^{\mathrm{eff}}$ we obtain from these combined data in Fig.~\ref{fig:sin2w}, in which we compare the extracted values of the weak mixing angle as obtained from $A_{LR}$ in isolation as well as from each of the leptonic asymmetries, $A_\ell$. We also obtain the globally fitted value of $\sin^2\theta_W^{\mathrm{eff}}$ which follows from jointly fitting these four channels.
In plotting Fig.~\ref{fig:sin2w}, we use our reconstructed SLD dataset discussed above ({\it i.e.}, the combined 1996--1998 SLD dataset).
This is consistent with the well-known fact that $A_{LR}$ provides the most precise SLD determination of $\sin^2\theta_W^{\mathrm{eff}}$, with the leptonic asymmetries giving lower-statistics but independent determinations.
Figure~\ref{fig:sin2w} also illustrates the consistency of the four independent determinations of $\sin^2\theta_W^{\mathrm{eff}}$ across the hadronic and leptonic channels.

\begin{table}[t]
\caption{Selected event counts and polarized electroweak observables extracted from the released 1996--1998 dataset.  $N_L$ ($N_R$) is the number of $Z\!\to\!q\bar{q}$ events recorded with left- (right-)polarized electron beam.  $\sin^2\theta_W^{\mathrm{eff}}$ and $\sin^2\theta_W^{\mathrm{lep}}$ are the effective weak mixing angles derived from $A_{LR}$ and from the universality-averaged leptonic asymmetry $A_\ell$, respectively.  The 1997 and 1998 runs are pooled into a single year-group following the published SLD analyses~\cite{SLD:2000leq,SLD:2000ujp}.  Uncertainties on the asymmetries are statistical and beam-polarization only.}
\label{tab:results}
\centering
\footnotesize
\setlength{\tabcolsep}{3pt}
\begin{tabular}{lccc}
\toprule
 & 1996 & 1997--1998 & Combined \\
\colrule
\multicolumn{4}{l}{\textit{Event counts}} \\
All reconstructed       & 102{,}699 & 555{,}542 & 658{,}241 \\
$Z\!\to\!q\bar{q}$      & 51{,}118  & 334{,}465 & 385{,}583 \\
$\;\;N_L$               & 28{,}584  & 184{,}894 & 213{,}478 \\
$\;\;N_R$               & 22{,}534  & 149{,}571 & 172{,}105 \\
$Z\!\to\!e^+e^-$        & 1{,}968   & 14{,}579  & 16{,}547 \\
$Z\!\to\!\mu^+\mu^-$    & 1{,}660   & 12{,}332  & 13{,}992 \\
$Z\!\to\!\tau^+\tau^-$  & 1{,}896   & 15{,}237  & 17{,}133 \\
\colrule
\multicolumn{4}{l}{\textit{Hadronic-channel asymmetry}} \\
$A_{LR}$                        & $0.1554{\pm}0.0065$ & $0.1445{\pm}0.0041$ & $0.1477{\pm}0.0035$ \\
$\sin^2\theta_W^{\mathrm{eff}}$ & $0.2305{\pm}0.0008$ & $0.2318{\pm}0.0005$ & $0.2314{\pm}0.0004$ \\
\colrule
\multicolumn{4}{l}{\textit{Leptonic-channel asymmetries}} \\
$A_e$    & $0.132{\pm}0.045$ & $0.165{\pm}0.016$ & $0.154{\pm}0.012$ \\
$A_\mu$  & $0.175{\pm}0.049$ & $0.130{\pm}0.018$ & $0.135{\pm}0.017$ \\
$A_\tau$ & $0.080{\pm}0.046$ & $0.116{\pm}0.016$ & $0.112{\pm}0.015$ \\
$A_\ell$ (univ.) & --- & --- & $0.137{\pm}0.008$ \\
$\sin^2\theta_W^{\mathrm{lep}}$ & --- & --- & $0.2328{\pm}0.0010$ \\
\botrule
\end{tabular}
\end{table}

\subsection{Scientific limitations}

These baselines should not be read as a re-derivation of the published SLD results.  Two limitations are worth highlighting explicitly.

\textit{Hadronic $A_{LR}$.}  Both our per-year-group hadronic event counts (and their $L/R$ splits) and the resulting raw $A_{LR}$ values track the corresponding numbers in the 2000 high-precision analysis~\cite{SLD:2000leq} very closely.  The residual differences are at the percent level on the counts and smaller than the polarization uncertainty on the asymmetry.  To compare with the Standard-Model prediction, however, the published analysis converts the measured $A_{LR}(E_\mathrm{cm})$ into the pole asymmetry $A_{LR}^0 = A_{LR}(E_\mathrm{cm}) + \Delta_\mathrm{EW}(E_\mathrm{cm})$, a small additive correction that accounts for pure-photon exchange, $\gamma$/$Z$ interference, and initial-state radiation, computed run-by-run with dedicated electroweak-fitting software (ZFITTER~6.22).  ZFITTER is no longer available in a form we can run on the released sample, so we report the raw, measured $A_{LR}$ at the dataset-averaged $E_\mathrm{cm}$ without this correction.  The resulting $\sin^2\theta_W^{\mathrm{eff}}$ therefore differs slightly from the published value by an amount consistent with the omitted shift, rather than a discrepancy in the underlying data.

\textit{Leptonic coupling asymmetries.}  Two effects complicate a direct numerical comparison with the 2001 leptonic-coupling paper~\cite{SLD:2000ujp}.  First, after applying our reconstruction of the published selection cuts, our selected event counts in the dilepton channels differ from those reported in 2001, most visibly in $\mu\mu$ and $\tau\tau$.  Because the original selection software is not available with the released data, the precise origin of this difference cannot be pinned down here.  It is most likely a residual difference in track-quality, particle-identification, or fiducial requirements that the public documentation does not fully specify.  Second, the published analysis uses an unbinned maximum-likelihood fit to the polarized differential cross-section with per-event acceptance corrections and a joint determination of $\mathcal{A}_e$ and $\mathcal{A}_\ell$, whereas we use direct event counting with an analytic fiducial correction $f_\mathrm{geom}(c)=(3+c^2)/(3c)$, where $c$ is the fiducial upper bound on $|\cos\theta_T|$, and take $\mathcal{A}_e$ from the muon-channel $L/R$ asymmetry.  Acceptance variation with $|\cos\theta_T|$ is therefore not corrected event-by-event in our extraction.  These two effects, together with the omission of all non-statistical systematic uncertainties, account for the modest pull of our universal $A_\ell$ relative to the 2001 published value.

Despite these caveats, the close per-year-group agreement of the hadronic counts and raw $A_{LR}$, the recovery of the expected $L/R$ splitting in all three leptonic $\cos\theta$ distributions, and the consistency of the four independently extracted $\sin^2\theta_W$ values support the central claim of this section: the translated dataset preserves the polarization-sensitive content of the SLD reconstruction at the level required for downstream physics and machine learning use, and is a faithful starting point for analyses that supply the missing radiative corrections and detector-level systematic treatment.

\section{Machine Learning Demonstration: Cross-Dataset Jet Embeddings}

We examine where SLD data sits in the latent space of a recent foundation model for particle physics.  Specifically, we embed SLD jets with the OmniLearned model~\cite{Bhimji:2025isp} alongside jets from widely used $e^+e^-$, $ep$, and $pp$ datasets and ask whether the SLD region of the latent space is qualitatively distinct from anything currently available to the public.  Of the four samples compared here, SLD is the only one consisting of reconstructed detector data, while the three comparison datasets are Monte Carlo simulations.  Because SLD events at the $Z$ pole have a back-to-back two-jet topology, rather than the single, narrow jet on which $pp$-trained models typically operate, we first cluster each event into two hemispheres with the $e^+e^-$ Durham $k_T$ algorithm~\cite{Catani:1991hj}, re-center each hemisphere on its jet axis, and embed the two hemispheres independently.

\begin{figure}[t]
\centering
\includegraphics[width=0.95\linewidth]{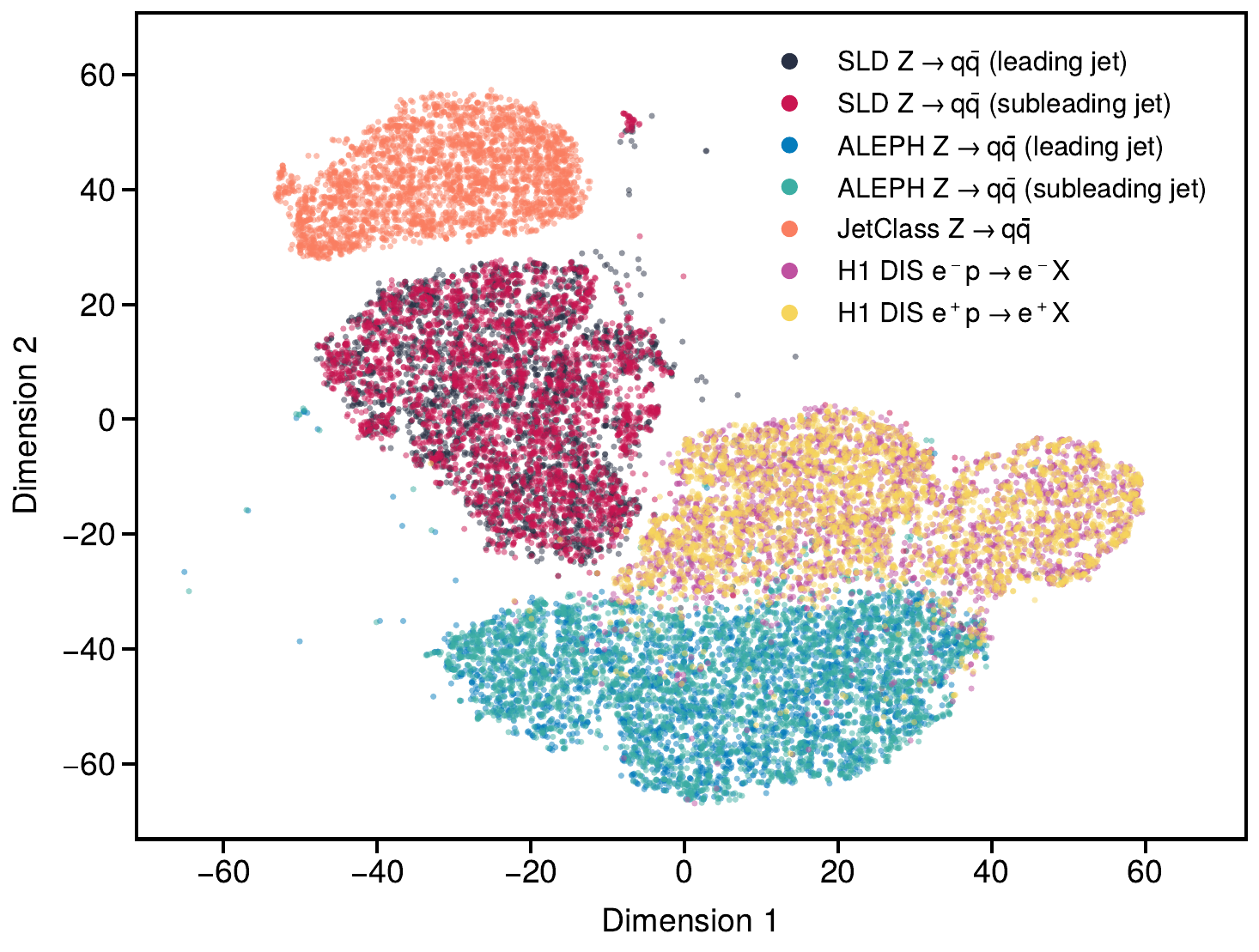}
\caption{t-SNE projection of OmniLearned jet embeddings across four datasets: SLD ($e^+e^-$, $Z$ pole, reconstructed detector data), ALEPH ($e^+e^-$, $Z$ pole)~\cite{Chen:2021aleph}, H1 ($ep$ DIS)~\cite{H1:2021data}, and JetClass ($pp$, LHC scale)~\cite{Qu:2022mxj}.  The latter three are Monte Carlo simulations.  Within each $Z\!\to\!q\bar{q}$ dataset the leading and subleading hemispheres collapse onto the same region, while the four datasets themselves form well-separated clusters that organize by initial state and energy scale.}
\label{fig:tsne}
\end{figure}

Projecting the resulting latent vectors into two dimensions via t-SNE~\cite{vanderMaaten:2008tsne} (Figure~\ref{fig:tsne}) reveals two patterns.  Within each $Z\!\to\!q\bar{q}$ dataset the leading and subleading hemispheres land on top of each other, as expected for two physically equivalent hemispheres.  The four datasets, by contrast, resolve into well-separated clusters: SLD and ALEPH share the same physics process but differ in detector and beam energies, while H1 ($ep$) and JetClass ($pp$) differ in both initial state and energy scale.  Because SLD is also the only reconstructed-data sample in this comparison, its cluster represents a regime, polarized $e^+e^-$ collisions at the $Z$ pole as actually recorded by the detector, that no current public dataset captures, and that is unavailable from simulation alone.  Releasing it therefore adds a genuinely new sample for ML research in HEP, whether for transfer-learning studies, domain-shift benchmarks, or training and evaluation on a regime current models have not seen.

\section{AI-Ready Documentation}\label{sec:documentation}

This section describes the documentation release introduced in Sec.~\ref{sec:dataset}: a digitized corpus of SLD-era internal notes, text extracted by four document-processing tools, and an agentic question-answering system built on top of the result. We evaluate the system end-to-end against a benchmark generated from the documents themselves. The scanned documents and extracted text are the durable release; the pipeline is supporting methodology. Because ground-truth transcriptions do not exist for this legacy corpus, standard Character Error Rate (CER) metrics are not available, and we instead compare extractor behavior through downstream performance on the agentic workflow---assessing whether the resulting text supports the LLM-driven tasks the release is intended to enable. The text-extraction stage includes one commercial system, the Azure AI Document Intelligence API, as a deliberate baseline against the three open-weight extractors. All components downstream of extraction---the reasoning agent, the
LLM-as-a-judge, the vision-language figure-description model, and the knowledge-graph builder---are designed to run on open-weight models served through the Argonne Leadership Computing Facility (ALCF) inference service. For the systematic validation studies reported in this section, Google models and endpoints were used as a higher-throughput fallback. The pipeline is intended to also work with open-weight models, keeping the agentic logic reproducible and free of commercial-API dependencies.

\subsection{The SLD-era internal-note corpus}
\label{sec:doc_corpus}

The corpus consists of approximately 1{,}190 documents from the SLD collaboration and the SLC accelerator program, scanned from physical copies in the SLAC archives\footnote{These documents were mostly borrowed from Martin Breidenbach, Tony Johnson, and Su Dong.}. Over 85\% date from 1980--1988, capturing the R\&D, design, and construction phases that preceded the high-statistics physics runs of the 1990s. The corpus is therefore mainly a record of how the experiment was built, rather than an operational manual.

The collection is dominated by SLC machine-physics notes---beam dynamics, wakefield mitigation, damping ring lattices, klystron timing---and SLD detector hardware specifications. Document quality is highly heterogeneous: many notes are second- or third-generation photocopies of typewritten originals, with hand-drawn figures, marginalia, and occasional handwritten equations alongside typeset ones. This makes the corpus a stress test for modern document-processing pipelines, which are typically benchmarked on born-digital or clean-scanned material.

\subsection{Text extraction}
\label{sec:doc_extraction}

We process the corpus with four tools: Marker~\cite{Paruchuri2023}, Docling~\cite{Aueretal2024}, Nougat~\cite{Blecheretal2023}, and the Azure AI Document Intelligence API~\cite{Microsoft2024}. Marker, Docling, and Nougat are open-weight models run locally on NERSC Perlmutter GPU nodes (each equipped with four NVIDIA A100 GPUs); Azure is included as a commercial baseline. Tools targeting born-digital PDFs are not applicable, since the documents are page images without an embedded text layer. The four extractors spans roughly an order of magnitude in wall time across the corpus---3.3\,h for Azure, 9.5\,h for Docling, 16.0\,h for Nougat (on 41\% coverage), and 23.7\,h for Marker---and a factor of two in mean output volume, from 2{,}151 tokens per document for Docling to 4{,}283 for Azure.

The extraction pipeline uses an SQL backend that tracks per-document provenance, tool versions, runtimes, and intermediate artifacts, so
that every released text file is traceable to a specific extractor
configuration. Text extraction itself dominates total compute at 97\% of wall time (52.5\,h); per-document summary and gist generation with Google's \texttt{gemini-3.1-flash-lite} accounts for 2.8\% (1.5\,h), and chunk embedding with \texttt{sentence-transformers/all-MiniLM-L6-v2}---a small, fast model not optimized for retrieval quality--for 0.3\% (8.9\,min). Parsing is the primary driver of compute cost in this pipeline.

\begin{figure*}[t]
    \centering
    \begin{subfigure}[b]{0.45\textwidth}
        \centering
        \includegraphics[width=\linewidth]{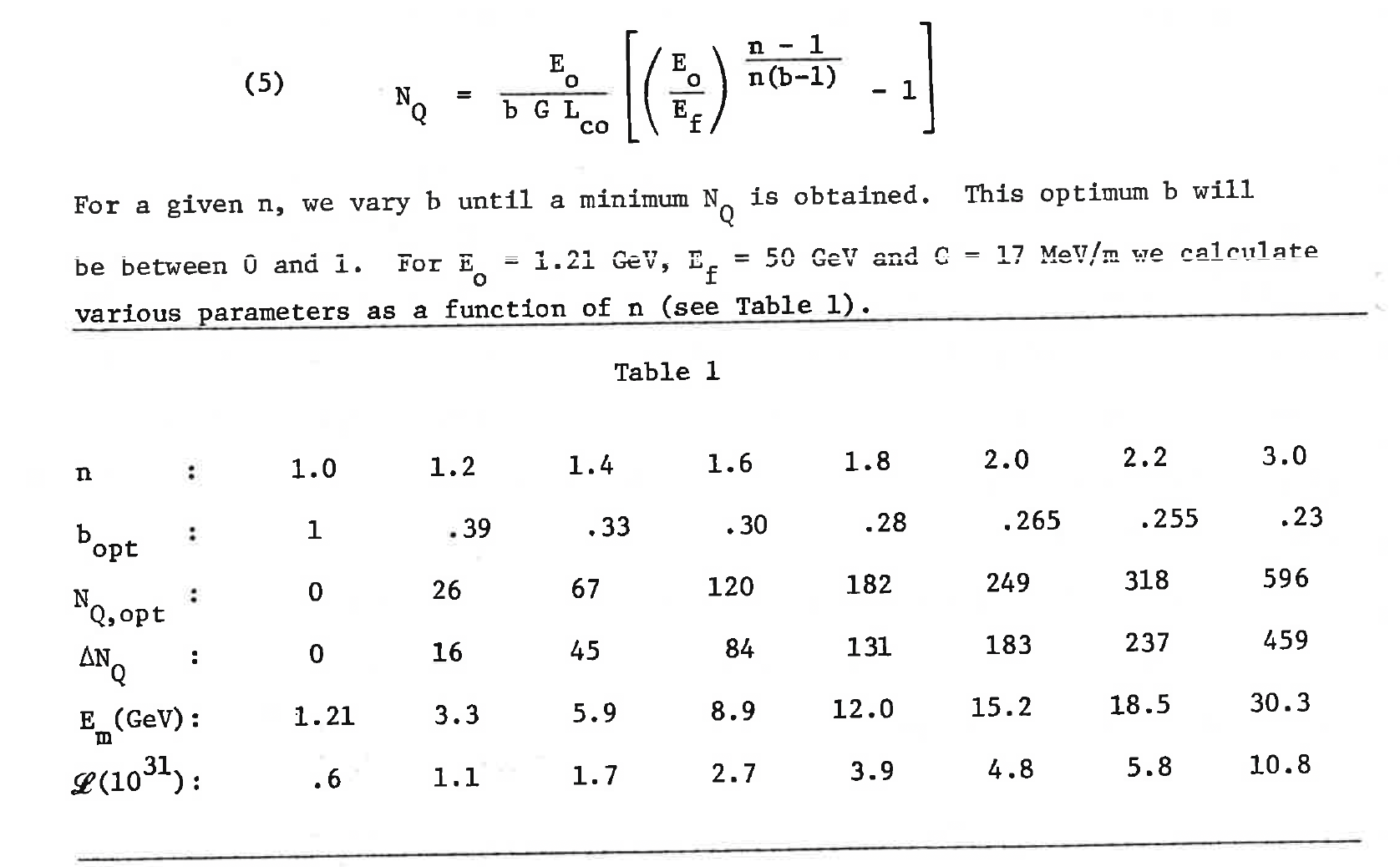}
        \caption{Example of complex data formatting. Marker recovers the block equation as valid \LaTeX\ but collapses the table; Azure produces the inverse.}
        \label{fig:doc:example1}
    \end{subfigure}
    \hfill
    \begin{subfigure}[b]{0.45\textwidth}
        \centering
        \includegraphics[width=\linewidth]{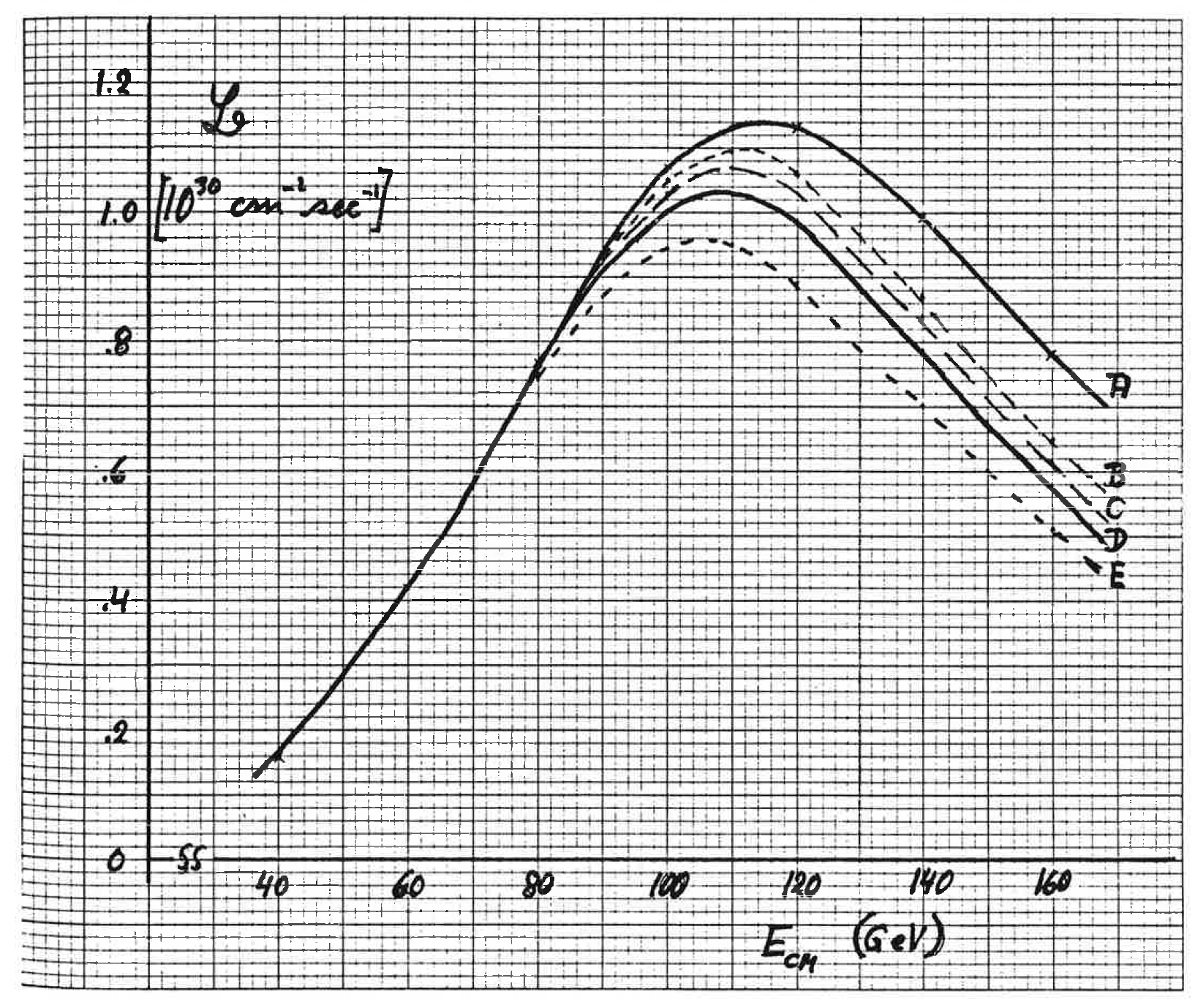}
        \caption{Example of hand-drawn figure. Azure OCRs gridlines into noise; Docling drops the image entirely; Marker plus VLM yields a semantic description.}
        \label{fig:doc:example2}
    \end{subfigure}
    \caption{Representative pages illustrating the distinct failure modes of the four extractors. Each panel pairs a scanned source page (left) with the extracted output discussed in the text.}
\end{figure*}

To compare extractors quantitatively beyond aggregate token counts, we evaluated parser outputs. For each document, an open-weight LLM produced a structured per-extractor assessment along five dimensions---equations, tables, figures, OCR quality, and structural preservation---categorizing each as clear success, partial success, or clear failure. The extractors specialize sharply along different axes: Marker excels at mathematical notation ($\sim$90\% clear success on equation-containing documents) but degrades on dense tables ($\sim$25\%), where its autoregressive decoder enters repetition loops that truncate the remainder of the document. Azure is the strongest at tabular data ($\sim$90\%), OCR ($\sim$90\%), and overall layout ($\sim$80\%), but degrades equations to flat ASCII in two-thirds of cases and occasionally misinterprets hand-drawn graph paper as enormous empty HTML tables. Figures see the largest spread: A VLM-augmented pipeline produces semantically rich descriptions in $\sim$90\% of figure-containing documents; Docling strips most figures with placeholders before reaching the description step. Nougat exhibits stability problems ($\sim$40\% repetition-loop or hallucinated-boilerplate failure rate on a document subsample) that make it unreliable for this corpus.


Figures~\ref{fig:doc:example1} and \ref{fig:doc:example2} make the same trade-offs concrete on two representative pages. The first is a typewritten note with a block equation and a dense parameter table. Marker translates the equation into valid \LaTeX\ but Azure degrades it to lossy ASCII and Docling strips it entirely, emitting a literal \texttt{<!-- formula-not-decoded -->} placeholder. For the table, Azure reconstructs valid HTML preserving alignment, while Marker collapses the columns into unreadable noise padded by repeated artifacts from autoregressive looping. The second example is a hand-drawn coordinate plot. Azure aggressively OCRs gridlines and axes, flooding the extraction with a huge, mostly empty HTML table; Docling identifies the region as an image but leaves only an empty placeholder. Combining Marker's figure extraction with a vision-language model bypasses raw OCR and replaces the image with a semantic description of the physics content (e.g., ``Line graph showing luminosity $L$ plotted against center-of-mass energy $E_{CM}$. Two distinct curves map the divergence with and without quadrupole wakefields''), which provides the downstream agent with searchable context that would otherwise be inaccessible to text-based retrieval.

\subsection{Demonstrating AI-readiness}
\label{sec:doc_demo}

We index each extractor's output and exercise it through an agentic question-answering system, generating an evaluation benchmark from the documents themselves and reporting performance across progressively more capable configurations. We emphasize that this is an illustration of AI-readiness tailored to scientific workflows, rather than a rigid NLP benchmark.

\textit{Indexing and retrieval.} For each extractor's output we run a per-document enrichment pass that produces a generated title, a one-paragraph gist, and a longer summary. Documents are chunked into overlapping token windows (1{,}000 tokens with 200-token overlap), and each chunk is prefixed with the document gist before being embedded with \texttt{sentence-transformers/all-MiniLM-L6-v2}, a small, fast model selected for throughput and not optimized yet for retrieval quality. Retrieval is hybrid: dense cosine similarity over chunk embeddings combined with PostgreSQL's native full-text search over the same chunks, with results merged into a single ranked list. We expose the corpus through several indices---chunk-level, gist-prefixed chunk,
and document-summary level.

\textit{Self-generated benchmark.} An LLM produces candidate question-answer pairs grounded in specific passages, prompted to paraphrase rather than copy terminology so that retrieval cannot succeed on lexical overlap alone. A LLM-audit pass discards pairs that are non-substantive, unanswerable from the cited passage, or dependent on copied wording. The benchmark inherits the corpus composition, with the majority of
questions concerning machine physics and detector hardware. From an
initial 657 candidate question-answer pairs, 463 (70\%) survived the LLM-judge audit and form the evaluation benchmark.

\textit{Performance across configurations.} We evaluate four configurations of increasing capability. A closed-book baseline using \texttt{gemini-3.1-pro-preview} answers $\sim$22\% of questions correctly, confirming that legacy SLD-era technical details are largely absent from frontier-model pretraining. Raw retrieval is also limited: hybrid embedding-and-keyword retrieval
at $K{=}10$ surfaces a passage from the source document in roughly
40\% of cases. This is partly a function of the small, untuned
embedding model, but it also reflects a structural feature of the
corpus---many topics (calibration constants, hardware revisions, the same subsystem at different design stages) appear across multiple notes, so the top-$K$ list is often dominated by topically relevant documents that are not the specific passage tagged as ground truth for a given question. We accordingly use an LLM-as-judge to evaluate task completion against the reference answer: a standard single-pass Retrieval-Augmented Generation (RAG) configuration---retrieve top-$K$ chunks, condition the generation on them---reaches 65\% task completion as graded by \texttt{gemini-3.5-flash}. The fully agentic configuration, in which the LLM iterates over a scratchpad and issues multiple targeted retrieval calls through a Model Context Protocol (MCP) server~\cite{MCP2024}, reaches near-saturation task completion (60/61) on a 61-question subset of the benchmark, evaluated at this scale because the multi-call agent's compute and token cost made full-benchmark runs impractical. The single-pass RAG baseline could likely be improved with retrieval tuning or chunking optimization; we report it as a reference point for what a non-iterative configuration achieves out of the box, not as a tuned competitor to the agent.

\textit{Why iteration matters.} The agent's advantage over single-pass RAG comes from query reformulation and multi-step retrieval rather than from any privileged information source. To illustrate, consider the question: 'Which CPU model and clock frequency are used in the tester's Microprocessor Module?' Single-pass RAG retrieves a structurally similar but incorrect document—a test driver board specification mentioning an 8748 microcontroller—and commits to that answer. The agent issues an initial search, attempts to fetch a candidate document and finds it does not match, reformulates the query to ``Intel 8085 Microprocessor,'' and on the third search retrieves the correct CTS Hybrid Tester System Configuration memorandum, from which it extracts the Intel 8085 at 6\,MHz and cites the source document.




\textit{Limitations.} Two caveats temper the near-saturation agent number. First, the benchmark is self-generated: An LLM writes questions grounded in passages, and an agent that searches passages is then evaluated on whether it surfaces those passages. The paraphrase requirement in question generation and the use of distinct generator and judge models partially mitigate this circularity, but the result should be read as ``the released corpus supports agentic exploration when the question is well-grounded in a single passage,'' not as a claim that the agent answers arbitrary external questions at 100\%. Second, LLM-judge evaluation is not infallible. On the question ``What is the physical location (Z coordinate relative to the start of SLC linac sector one) of the Ring-To-Linac (RTL) and Linac interface?,'' the benchmark expected $Z = 101.6000$\,m, while the agent retrieved the source memo and answered $Z = 98.7532$\,m, identifying this as the ``treaty point'' between Sector~1, the LTR/RTL transport lines, and Sector~2. Both numbers appear in the source document and the agent's answer is arguably the more defensible interpretation of the question, but the judge marked it incorrect because it did not match the reference string. Cases like this contribute a small but non-zero source of judge error in the reported numbers.


\section{Conclusions and Outlook}

We have presented a modernized, publicly available dataset from the SLD experiment at the SLAC Linear Collider.  The dataset comprises approximately 660{,}000 reconstructed events from the 1996--1998 running of the experiment, recorded with a highly polarized electron beam, converted from legacy formats and accompanied by digitized internal documentation.
We additionally distribute the open-source \texttt{jazelle} translation toolkit, which can serialize the release into HDF5, Parquet, or Feather as well, and we have validated the translation by reproducing the canonical SLD electroweak measurements ($A_{LR}$ and $A_\ell$) on the full released 1996--1998 dataset.

This release serves several purposes.  First, it preserves an important dataset from a unique collider that will not be replicated in the foreseeable future---SLD remains the only experiment to have collected a large sample of $Z$ decays with polarized beams.  Second, it provides a valuable testbed for developing and benchmarking ML techniques in particle physics: the clean $e^+e^-$ environment, precisely known initial state, and unique polarization information offer a complementary setting to the hadron collider datasets that dominate current ML research.  Third, the dataset may enable new physics analyses that were not envisioned during the original operation of SLD, leveraging advances in both ML methods and theoretical understanding.

The cross-dataset embedding study with OmniLearned makes the second point concrete: SLD jets occupy a region of the foundation model's latent space that is well separated from those of ALEPH, H1, and JetClass.  As SLD is the only reconstructed-data sample in that comparison, this also identifies a corner of the experimental landscape, polarized $e^+e^-$ collisions at the $Z$ pole in real detector data, that is not currently represented in public collider datasets used by the ML community.

Looking ahead, the dataset could be extended with Monte Carlo simulations to enable detector-level studies, and could serve as a prototype for similar preservation and modernization efforts at other legacy experiments.  We encourage the community to explore this dataset and welcome contributions to expand its scope and utility.

\section*{Acknowledgments}

We are grateful to the SLAC accelerator staff that operated SLC and the SLD Collaboration for collecting and curating this unique and scientifically rich dataset.  These activities were supported by the U.S. Department of Energy (DOE) and the U.S. National Science Foundation; the Istituto Nazionale di Fisica Nucleare of Italy; the Japan–US Cooperative Research Project on High Energy Physics; and, in the United Kingdom, the Science and Engineering Research Council and its successor the Particle Physics and Astronomy Research Council.

We thank the SLD Collaboration for allowing us to resurrect and release their data and documentation; in particular, we are indebted to Martin Breidenbach, Tony Johnson, and Su Dong for extensive conversations and for loaning us artifacts related to these data.

We also thank Luna Chen and our other colleagues in the electron-positron alliance for useful conversations.

This work was supported by the DOE Office of Science under Contracts DE-AC02-06CH11357 and DE-AC02-76SF00515.  In particular, this research was catalyzed and carried out by the Genesis Mission, through the HEP pilot project Knowledge Extraction.  We are grateful to Katrin Heitmann and all of our other collaborators on this project.

\section*{Code and data availability}

The dataset is available on Zenodo at \url{https://zenodo.org/records/19925960}.  The accompanying Python translation toolkit, \texttt{jazelle}, is available at \url{https://github.com/HEP-KE/jazelle_reader}, and the analysis code used to reproduce the measurements presented here is available at \url{https://github.com/HEP-KE/sld-resurrect}.

\bibliography{main}
\bibliographystyle{JHEP}

\end{document}